\input{epsf}
\documentclass[11pt]{article}
\def\be{\begin{equation}}
\def\eea{\end{eqnarray}}
\def\bea{\begin{eqnarray}}
\def\ee{\end{equation}}

\author{M.Mohammadi$^{1,2}$ \footnote{majid471702@yahoo.com} , M.H.Naderi$^{3}$ \footnote{mhnaderi2001@yahoo.com} and M.Soltanolkotabi$^{3}$ \footnote{soltan@sci.ui.ac.ir}
\\ $^{1}$ {\small Physics Department, Science and Research Campus Azad University Tehran, Iran.}
\\$^{2}$ {\small Physics Department, Shahreza and Islamic Azad University, Shahreza, Isfahan, Iran.}
\\$^{3}$ {\small Quantum Optics Group, University of Isfahan, Isfahan, Iran.}}
\title{Influence of gravitational field on quantum-nondemolition
          measurement of atomic momentum in the dispersive Jaynes-Cummings
           model}
 \begin{document}
\maketitle
\begin{abstract}
\noindent We present a theoretical scheme based on su(2) algebra
to investigate the influence of homogeneous gravitational field on
the quantum nondemolition measurement of atomic momentum in
dispersive Jaynes-Cummings model. In the dispersive
Jaynes-Cummings model, when detuning is large and the atomic
motion is in a propagating light wave, we consider a two-level
atom with quantized cavity-field in the presence of a homogeneous
gravitational field. We derive an effective Hamiltonian
describing the dispersive atom-field interaction in the presence
of gravitational field. We can see gravitational influence both on
the momentum filter and momentum distribution. Moreover,
gravitational field decreases both tooth spacing of momentum and
the width of teeth of momentum.
\end{abstract}
\\
\section{Introduction}
Among the models describing the interaction between light and
matter, the Jaynes-Cummings model (JCM) [1] seems to be ideal. The
JCM describes the interaction between a two-level atom and a
single quantized mode of the electromagnetic field in a lossless
cavity within the rotating wave approximation (RWA). This is the
simplest model of the radiation-matter interaction. It is simple
enough to be exactly solved on the one hand and complicated
enough to exhibit many fascinating quantum features on the other
hand. These pure quantum effects include quantum collapses and
revivals of atomic inversion [2], squeezing of the radiation field
[3] ,atomic dipole squeezing [4], vacuum Rabi oscillation [5] and
dynamical entangling and disentangling of the atom field system
in the course of time [6-8]. Further interest the JCM comes from
the fact that its theoretical predictions have been extensively
used in the context of quantum information [9], atoms and ions
trapping [10,11] and quantum nondemolition (QND) measurements [12].\\
\hspace*{00.5 cm} In a general QND measurement, an observable
signal of a quantum system is measured by detecting a change in an
observable of the probe system coupled to the quantum system
during the measurement time, without perturbing the subsequent
evolution of the observable signal. We can therefore make a
sequence of precise measurements of an observable signal such
that the result of each measurement is completely predictable
from the result of the preceding measurement. Such an observable
is called QND observable. Original QND ideas involved a
dispersive coupling of the signal field to a material probe [13].
The QND method is quite generally based on dispersive and
nonlinear effects. Methods to avoid the back action of the
measurement on the detected observable have been proposed and
implemented in the optical domain [14,15]. These experiments are
the realization of the QND schemes introduced in [13]. They rely
on nonlinear coupling of the signal field to be measured with a
probe field whose phase is altered by a quantity depending on the
number of photons in the signal beam. In a paper by Sleator and
Wilkens [16] it was shown a complementary scheme in which a
quadrature-component of a propagating laser wave acts as the
probe for the QND measurement of the atomic momentum. It is based
on the Doppler effect on the component of atomic momentum along
the propagation direction of the light field.\\
\hspace*{00.5 cm} On the other hand, experimentally, atomic beams
with very low velocities are generated in laser cooling and
atomic interferometry [17]. It is obvious that for atoms moving
with a velocity of a few centimeters or meters per second for a
time period of several milliseconds or more, the influence of the
Earth's acceleration becomes important and cannot be neglected
[18]. For this reason it is of interest to study the temporal
evolution of an atom moving in the gravitational field of the
Earth and the cavity-field. Since any optical experiment in the
laboratory is in a non-inertial frame it is important to estimate
the influence of the Earth's acceleration on the outcome of these
experiments. The effective Hamiltonian and time evolution for
two-level atom that is simultaneously exposed to the field of a
running laser wave and a homogeneous gravitational field is
studied by Marzlin [19].\\
\hspace*{00.5 cm} In this paper we investigate a complementary
scheme based on su(2) algebra [20] to investigate the influence
of gravitation on the QND measurement of atomic momentum in the
dispersive JCM. In section 2 we apply scheme with su(2) algebra
to describe the interaction of a two-level atom with quantized
field and a homogeneous gravitational field in dispersive JCM.
Recently, optical Schr\"{o}dinger-cat states have been realized
in dispersive JCM [21]. Also, these states have been verified
experimentally, by Auffeves and coworkers, for a two-level atom
interacting with a single mode of the electromagnetic field in
dispersive JCM [22]. In the dispersive JCM, the atom is in ground
state and detuning is large, so one can neglect spontaneous
emission. In section 3 we investigate dynamical evolution of the
system and show that how the gravitational field may be affected
the dynamical properties of the dispersive JCM. In section 4 we
investigate the influence of gravitational field on the QND
measurement. Finally, we summarize our conclusions in section 5.
\section{Dispersive Jaynes-Cummings Model in the presence of Gravitational Field}
In the dispersive JCM, we assume that the atom is in its ground
state initially and we consider the case of large detuning so
that the excited state of the atom is almost never populated, so
we neglect spontaneous emission. The importance of dispersive JCM
is because of using in generation schr\"{o}dinger cat states with
superposition of coherent states. These states have been
generated in different contexts. A great variety of methods have
been proposed for generate such states, for example, in a
Mach-Zehnder interferometer [23] and one of the first schemes due
to Yurke and Stoler [24] who showed that a coherent state(CS)
propagating in a Kerr medium could lead to a schr\"{o}dinger cat
state.  We take into account the atomic motion along the position
vector $\hat{\vec{x}}$, so the evolution of the atom-field system
in the presence of gravitational field  is governed by the
Hamiltonian
\begin{eqnarray}
\hat{H}=&&\frac{\hat{p}^{2}}{2M}-M\vec{g}.\hat{\vec{x}}+\hbar\omega_{c}(\hat{a}^{\dag}\hat{a}+\frac{1}{2})+\frac{1}{2}\hbar\omega_{eg}\hat{\sigma}_{z}+\nonumber\\
&&\hbar\lambda[\exp(-i\vec{q}.\hat{\vec{x}})\hat{a}^{\dag}\hat{\sigma}_{-}+\exp(i\vec{q}.\hat{\vec{x}})\hat{\sigma}_{+}\hat{a}],
\end{eqnarray}
where $\hat{a}$ and $\hat{a}^{\dag}$ denote, respectively, the
annihilation and creation operators of a single-mode traveling
wave with frequency $\omega_{c}$, $\vec{q}$ is the wave vector of
the running wave and $\hat{\sigma}_{\pm}$ denote the raising and
lowering operators of the two-level atom with electronic levels
$|e\rangle, |g\rangle $ and Bohr transition frequency
$\omega_{eg}$. The atom-field coupling is given by $\lambda$ and
 $\hat{\vec{p}}$, $\hat{\vec{x}}$
denote, respectively, the momentum and position operators of the
atomic center of mass motion and $g$ is the Earth's gravitational
acceleration. We find that an alternative representation of su(2)
algebra arises naturally from the system.We construct a
representation of su(2) algebra based on the generalized algebra
and the pauli matrices. From Hamiltonian (1), it is apparent that
there exist an operator $\hat{K}$ with constant of motion
\begin{equation}
\hat{K}=\hat{a}^{\dag}\hat{a}+|e\rangle \langle e|.
\end{equation}
In addition, the operator
$\hat{a}\hat{\sigma_{+}}=\hat{a}|e\rangle \langle g|$ commutes
with $\hat{K}$. Now we introduce the following operators
\begin{equation}
\hat{S_{0}}=\frac{1}{2}(|e \rangle \langle e|-|g \rangle \langle
g|) , \hat{S_{+}}=\hat{a}|e\rangle \langle
g|\frac{1}{\sqrt{\hat{K}}},
\hat{S_{-}}=\frac{1}{\sqrt{\hat{K}}}|g\rangle \langle
e|\hat{a}^{\dag}.
\end{equation}
We can show that the operators $\hat{S_{0}}$, $\hat{S_{\pm}}$
satisfy the following commutation relations
\begin{equation}
[\hat{S_{0}},\hat{S_{\pm}}]=\pm
\hat{S_{\pm}},[\hat{S_{-}},\hat{S_{+}}]=-2\hat{S_{0}},
\end{equation}
where $S_{0}$, $S_{\pm}$ are the generators of su(2) algebra [25].
In terms of su(2) generators, the Hamiltonian(1) can be rewritten
as
\begin{equation}
\hat{H}=\frac{\hat{p}^{2}}{2M}-M\vec{g}.\hat{\vec{x}}+\hbar
\omega_{c}\hat{K}+\frac{1}{2}\hbar\triangle\hat{S_{0}}+\hbar\lambda\sqrt{\hat{K}}(\exp(-i\vec{q}.\hat{\vec{x}})\hat{S_{-}}+\exp(
i\vec{q}.\hat{\vec{x}} )\hat{S}_{+}),
\end{equation}
where
\begin{equation}
\triangle=\omega_{eg}-\omega_{c},
\end{equation}
is the usual detuning parameter.\\
 Now we start to find the exact solution for the dynamical evolution of the total system governed
by the Hamiltonian(5). The corresponding time evolution operator
can be expressed as
\begin{equation}
\hat{u}(t)=\exp(\frac{iM\vec{g}.\hat{\vec{x}}t}{\hbar})\hat{v}^{\dag}\hat{u}_{e}(t)\hat{v},
\end{equation}
where
\begin{equation}
\hat{v}=\exp(-i\vec{q}.\hat{\vec{x}}\hat{S}_{0}),
\end{equation}
\begin{equation}
\hat{u}_{e}=\exp(\frac{-i\hat{H}_{e}t}{\hbar}).
\end{equation}
It can be shown that the operator $\hat{u}_{e}(t)$ satisfy an
effective Schr$\ddot{o}$dinger equation governed by an effective
Hamiltonian $\hat{H}_{e}$, that is
\begin{equation}
i\hbar\frac{\partial\hat{ u}_{e}}{\partial
t}=\hat{H}_{e}\hat{u}_{e},
\end{equation}
where
\begin{equation}
\hat{H}_{e}=\frac{\hat{p}^{2}}{2M}-\hbar
\hat{\triangle}(\hat{\vec{p}},\vec{g})\hat{S}_{0}+\frac{1}{2}Mg^{2}t^{2}+\vec{g}.\hat{\vec{p}}t+\hbar
\lambda
(\sqrt{\hat{K}}\hat{S}_{-}+\sqrt{\hat{K}}\hat{S}_{+})+\hat{H}_{0},
\end{equation}
\begin{equation}
\hat{H}_{0}=\hbar \omega_{c}\hat{K}-\frac{\hbar}{2}\triangle
\hat{S}_{0}-\frac{q^{2}\hbar^{2}}{2M}\hat{S}_{0}+\frac{q^{2}\hbar^{2}}{8M},
\end{equation}
and the operator
\begin{equation}
\hat{\triangle}(\hat{\vec{p}},\vec{g})=\omega_{c}-(\omega_{eg}+\frac{\vec{q}.\hat{\vec{p}}}{M}+\vec{q}.\vec{g}t+\frac{q^{2}\hbar}{2M}),
\end{equation}
has been introduced as the Doppler shift detuning at time t.
Therefore, due to the Doppler shift of
$\frac{\vec{q}.\hat{\vec{p}}}{M}$ , recoil frequency
$\omega_{rec}=\frac{\hbar q^{2}}{2M}$ and gravitational field, the
detuning between the cavity and the atomic transition frequency
has been modified. The relevant time scale introduced by the
gravitational influence is
\begin{equation}
\tau_{a}=\frac{1}{\sqrt{\vec{q}.\vec{g}}}.
\end{equation}
For an optical laser with $(|\vec{q}|\simeq 10^{7}m^{-1})$ and
Earth's acceleration $(|\vec{g}|=9.8\frac{m}{s^{2}})$, $\tau_{a}$
is about $10^{-4}s$. We remark that
$\hat{\triangle}(\hat{\vec{p}},\vec{g})$ does only depend on the
product $\vec{q}.\vec{g}$, so that the influence of the
gravitational acceleration on the internal evolution vanishes if
the acceleration is perpendicular to the laser field.
 Now we apply the interaction picture, i.e.,
\begin{equation}
\hat{u}_{e}=\exp(\frac{-it\hat{H}_{0}}{\hbar})\hat{\tilde{u}},
\end{equation}
such that
\begin{equation}
i\hbar\frac{\partial\hat{\tilde{u}}}{\partial
t}=\hat{\tilde{H}}\hat{\tilde{u}},
\end{equation}
where
\begin{equation}
\hat{\tilde{H}}=\frac{\hat{p}^{2}}{2M}-\hbar
\hat{\triangle}(\hat{\vec{p}},\vec{g})\hat{S}_{0}+\frac{1}{2}Mg^{2}t^{2}+\hat{\vec{p}}.\vec{g}
t+\hbar (\kappa
\sqrt{\hat{K}}\hat{S}_{-}+\kappa^{*}\sqrt{\hat{K}}\hat{S}_{+}),
\end{equation}
and $\kappa(t)$ is an effective coupling coefficient
\begin{equation}
\kappa= \lambda
\exp(\frac{it}{2}(\hat{\triangle}(\hat{\vec{p}},\vec{g})+\frac{\hbar
q^{2} }{M})).
\end{equation}
In the limit of very small values of
$|\langle\hat{\triangle}(\hat{\vec{p}},\vec{g})\rangle|$ and
$\frac{\hbar q^{2}}{M}$, the coefficient $\kappa(t)$ is
independent of time. As it stands, the effective Hamiltonian (17)
has the form of the Hamiltonian of the JCM, the only modification
being the dependence of the detuning on the conjugate momentum
and the gravitational field . Now we use the JCM in the dispersive
limit. In this limit, we assume that the atom is in its ground
state initially and we consider the case of large detuning,
$|\delta|\rangle \rangle \kappa \sqrt{\langle
\hat{a}^{\dag}\hat{a}\rangle}$, with
$\delta\equiv\omega_{c}-\omega_{eg}-\omega_{rec}$. In this case,
the excited state of the atom is almost never populated, so we
neglect spontaneous emission. Now we obtain effective Hamiltonian
in dispersive regime. In the interaction picture the transformed
Hamiltonian (17) takes the following form
\begin{equation}
\hat{\tilde{H}}_{int}=\exp(\frac{-i\hat{\tilde{H}}_{0}t}{\hbar})\hat{\tilde{H}}_{I}\exp(\frac{i\hat{\tilde{H}}_{0}t}{\hbar}),
\end{equation}
where
\begin{equation}
\hat{\tilde{H}}_{0}=-\hbar
\hat{\triangle}(\hat{\vec{p}},\vec{g})\hat{S}_{0},
\end{equation}
\begin{equation}
\hat{\tilde{H}}_{I}=\hbar(\kappa
\sqrt{\hat{K}}\hat{S}_{-}+\kappa^{*}\sqrt{\hat{K}}\hat{S}_{+})+\hat{H}(\hat{\vec{p}},\vec{g}),
\end{equation}
\begin{equation}
\hat{H}(\hat{\vec{p}},\vec{g})=\frac{\hat{p}^{2}}{2M}+\hat{\vec{p}}.\vec{g}t+\frac{1}{2}Mg^{2}t^{2}.
\end{equation}
Therefore we obtain
\begin{equation}
\hat{\tilde{H}}_{int}=\hbar (\kappa \sqrt{\hat{K}}\hat{S}_{-}
\exp(-it\hat{\triangle}(\hat{\vec{p}},\vec{g}))+\kappa^{*}\sqrt{\hat{K}}\hat{S}_{+}\exp(it\hat{\triangle}(\hat{\vec{p}},\vec{g})))+\hat{H}(\hat{\vec{p}},\vec{g}).
\end{equation}
 We can obtain the effective Hamiltonian in dispersive limit [26]
\begin{equation}
\hat{H}_{eff}=\hat{H}(\hat{\vec{p}},\vec{g})+\hbar
\Omega(\hat{\vec{p}},\vec{g})\hat{a}^{\dag}\hat{a},
\end{equation}
where
\begin{equation}
\hat{\Omega}(\hat{\vec{p}},\vec{g})=\frac{|\kappa|^{2}}{\hat{\triangle}(\hat{\vec{p}},\vec{g})},
\end{equation}
is the momentum-dependent frequency of the harmonic oscillator
and identified as the Doppler modified ac stark shift of the
atom-field interaction.
\section{Dynamical Evolution}
In section 2, we obtained the effective Hamiltonian atom-field
system in the presence of the gravitational field in dispersive
regime. In this section, we investigate dynamical evolution of
system. We will show how the gravitational field may be affected
the dispersive JCM. We will also investigate the dispersive JCM,
in the short and long time limits. The Schr\"{o}dinger equation
reads as
\begin{equation}
i \hbar \frac{\partial |\psi \rangle } {\partial
 t}=\hat{H}_{eff}|\psi\rangle ,
\end{equation}
where
\begin{equation}
|\psi(t) \rangle =|\psi_{g}(t)\rangle \otimes |g\rangle.
\end{equation}
In the dispersive regime, we define $|\psi_{g}(t)\rangle $ as the
state of the center of mass and the cavity field. We assume at
$t=0$, the atom-field system is described by the product state
where the cavity field is initially prepared in a coherent state
$|\alpha \rangle$. We apply evolution operator
\begin{equation}
\hat{u}(t)=\exp(\frac{-i}{\hbar}\int_{0}^{t}\hat{H}_{eff}(t')dt'),
\end{equation}
on the initial state
\begin{equation}
|\psi_{g}(t=0)\rangle=(\int d^{3}p
\phi_{g}(\vec{p})|\vec{p}\rangle)\otimes|\alpha\rangle,
\end{equation}
where $\phi_{g}(\vec{p})$ is the probability amplitude for the
center-of-mass motion of the ground-state atom in the momentum
representation,
 $\hat{\vec{p}}|\vec{p}\rangle=\vec{p}|\vec{p}\rangle$. When the
atom leaves the cavity after an interaction time $\tau$, the state
vector has evolved into the entangled state
\begin{equation}
|\psi_{g}(t=\tau)\rangle=\hat{u}(t=\tau)|\psi_{g}(t=0)\rangle,
\end{equation}
\begin{eqnarray}
|\psi_{g}(t=\tau)\rangle=&&\int d^{3}p \exp(\frac{-i\tau
p^{2}}{2M\hbar})\exp(\frac{-i\vec{p}.\vec{g}\tau^{2}}{2\hbar})\exp(\frac{-iMg^{2}\tau^{3}}{6\hbar})\phi_{g}(\vec{p})|\vec{p}\rangle\nonumber\\&&\otimes|\alpha\exp(-i\Omega(\vec{p},\vec{g})\tau)\rangle.
\end{eqnarray}
We now consider gravitational influence on the dynamical
evolution of the system for two limiting case. The first, in the
limit of small gravitational influence, $t<<\tau_{a}$, means very
small $\vec{q}.\vec{g}$, i.e., the momentum transfer from the
laser beam to the atom is only slightly altered by the
gravitational acceleration because the latter is very small or
nearly perpendicular to the laser beam. In this limit, the state
vector is
\begin{equation}
|\psi_{g}(t=\tau)\rangle=\int d^{3}p \exp(\frac{-i\tau
p^{2}}{2M\hbar})\phi_{g}(\vec{p})|\vec{p}\rangle\otimes|\alpha\exp(-i\Omega(\vec{p})\tau)\rangle,
\end{equation}
where
\begin{equation}
\Omega(\vec{p})=\frac{|\hat{\kappa}|^{2}}{\triangle(\vec{p})},\triangle(\vec{p})=\omega_{c}-(\omega_{eg}+\frac{\vec{q}.\vec{p}}{M}+\frac{q^{2}\hbar}{2M}).
\end{equation}
The Doppler shift detuning is independent of gravitational field.
The second, in the limit of long times, $t>>\tau_{a}$, the atoms
are accelerated by the Earth's gravity so that their velocity
increases and the Doppler shift detuning in (13) is depends on
the gravitational field.
\section{ The QND Measurement}
The QND measurement has been the subject of numerous
investigations in the past two decades [27-30]. For such a
measurement on of a system, the system must be coupled to another
system (called probe), and monitored an appropriately selected
probe observable during the measurement. The system-probe
interaction has to be chosen in such a way that the corresponding
interaction Hamilton commutes with the system observable. The
interaction of radiation with a single atom involves the
electronic degrees of freedom of atom and center-of-mass degrees
of freedom. The interaction of a two-level atom with a standing
laser wave can result in a QND measurement of the atomic position
when a quadrature component of the (sufficiently detuned) laser
field is used as the probe [31,32]. In the previous section we
show that how the gravitational field could be affected the
dispersive Jaynes-Cummings model. In this section we introduce
QND measurement and investigate the influence of gravitational
field on the QND measurement of atomic motion. The Hermitian
quadrature phase operator
$\hat{Y}=\frac{(\hat{a}+\hat{a}^{\dag})}{2}$ is used as a probe
observable for a QND measurement of conjugate momentum
$\hat{\vec{p}}$ when (17) satisfies  condition
$\hat{\tilde{H}}=\hat{\tilde{H}}(\hat{\vec{p}}),[\hat{\tilde{H}},\hat{\vec{p}}]=0,[\hat{\tilde{H}},\hat{Y}]\neq
0$. The probability for obtaining value $Y$ for the quadrature
phase $\hat{Y}$ may be expressed as
\begin{equation}
P(Y)dY=dY\int dp |\phi_{g}(\vec{p})|^{2} |\langle
Y|\alpha\exp(-i\Omega(\vec{p},\vec{g})\tau)\rangle|^{2}.
\end{equation}
The momentum distribution after a readout Y is given by the
conditional probability $P(\vec{p}|Y)$, that the atom has a
momentum vector $\vec{p}$
\begin{equation}
P(\vec{p}|Y)=|\phi_{g}(\vec{p})|^{2} |\langle
Y|\alpha\exp(-i\Omega(\vec{p},\vec{g})\tau)\rangle|^{2}P(Y)^{-1},
\end{equation}
where
\begin{eqnarray}
\langle
Y|\alpha\exp(-i\Omega(\vec{p},\vec{g})\tau)\rangle=&&(\frac{2}{\pi})^{\frac{1}{4}}\exp(-[|\alpha|\cos(\Omega(\vec{p}.\vec{g})\tau
-\varphi_{\alpha})-Y]^{2})-\nonumber\\&&2i|\alpha|Y\sin(\Omega(\vec{p},\vec{g})\tau-\varphi_{\alpha}).
\end{eqnarray}
We define the momentum filter $G(\vec{p})=|\langle
Y|\alpha\exp(-i\Omega(\vec{p},\vec{g})\tau)\rangle|^{2}$. The
readout $Y$ implies that the atom has a momentum that obeys
$\cos[\Omega(\vec{p},\vec{g})\tau-\varphi_{\alpha}]\simeq
\frac{Y}{|\alpha|}$, with
$\alpha=|\alpha|\exp(i\varphi_{\alpha})$. Figures 1a and 2a show
the form of the momentum filter $|\langle
Y|\alpha\exp(-i\Omega(\vec{p},\vec{g})\tau)\rangle|^{2}$ and the
momentum distribution $P(\vec{p}|Y=0)$ respectively assuming
$Y=0$. In these figures we assume
$(\frac{\kappa}{\delta})^{2}\tau\omega_{rec}=0.2,
\tau\omega_{rec}=7.2, \kappa \tau=140, \alpha=2, q=10^{7}m^{-1},
M=10^{-26}kg, g=9.8\frac{m}{s^{2}}, \tau=14.4\times 10^{-6}s$, and
$\varphi_{\alpha}=\Omega(0)\tau+\frac{\pi}{2}$ [11,20]. Here we
consider a beam of two-level atoms traversing an arm of an
optical ring cavity, so that $\vec{p}.\vec{q}=pqCos\theta,
\vec{q}.\vec{g}=qgSin\theta,$ with $\theta=\frac{\pi}{4}$. In
figures 1a-1d and 2a-2d we plot the momentum filter and momentum
distribution in terms of $\frac{p}{\hbar q}$. In these figures we
show that gravitational field influence on the momentum filter and
momentum distribution when the time increases. In figures 2a-2d
one can see Oscillations. These oscilations result from quantum
interference of translation motion [21]. To estimate the spacing
for small $\vec{p}$, we expand $\Omega(\vec{p})\simeq
\Omega(0)+(\frac{\kappa}{\delta-\vec{q}.\vec{g}\tau})^{2}\frac{\vec{q}.\vec{p}}{M}
$ and obtain
\begin{equation}
\triangle \vec{p}=|\vec{p}_{n+1}-\vec{p}_{n}|=\hbar q
\frac{\pi}{2}(\frac{\delta-\vec{q}.\vec{g}\tau}{\kappa})^{2}\frac{1}{\omega_{rec}\tau}.
\end{equation}
The slow variation of $\triangle \vec{p}$ in figures 1a-1d is due
to the nonlinearity of $\Omega(\vec{p},\vec{g})$, which leads to a
decreasing tooth spacing for increasing momenta. In a simple
Gaussian approximation, the width of the teeth near $p=0$ is
given by
\begin{equation}
\sigma=\hbar q
\frac{1}{4|\alpha|}(\frac{\delta-\vec{q}.\vec{g}\tau}{\kappa})^{2}\frac{1}{\omega_{rec}\tau}.
\end{equation}
From (37) and (38) one can see that the gravitational field
decreasing both $\triangle \vec{p}$ and $\sigma$.
\section{Summary and conclusions}
In this paper we have investigated the influence of gravitational
field on the dynamical behavior of the JCM and on the QND
measurement of atomic momentum in dispersive JCM. For this
purpose, based on su(2) algebra as the dynamical symmetry group
of the model, we have derived an effective Hamiltonian describing
the dispersive atom-field interaction in the presence of
gravitational field. By finding an explicit form for the
corresponding time evolution operator, we have explored the
influence of gravitation on the atom-field coupling and detuning
parameter. We have shown that due to the gravitational field the
atomic transition frequency experiences a Doppler shift and
atom-field coupling becomes time-dependent. Then we have
investigated the influence of gravitational field on the QND
measurement of atomic momentum in dispersive JCM. Moreover, we
have shown that the gravitational field decreases both tooth
spacing of momentum and the width of teeth of momentum.\\
\\
{\bf  Acknowledgements} \\
On of the authors (M.M) wishes to thank
The Office of Graduate Studies of the Science and Research Campus
Islamic Azad University of Tehran for their support.

\vspace{20cm}

{\bf FIGURE CAPTIONS:}

{\bf FIG. 1 } The momentum filter $G(\vec{p})=|\langle
Y|\alpha\exp(-i\Omega(\vec{p},\vec{g})\tau)\rangle|^{2}$ as a
function of $\frac{p}{\hbar q}$ that results a readout $Y=0$.
Here $(\frac{\kappa}{\delta})^{2}\tau\omega_{rec}=0.2,
\tau\omega_{rec}=7.2, \kappa \tau=140, \alpha=2, q=10^{7}m^{-1},
M=10^{-26}kg, g=9.8\frac{m}{s^{2}}$.

{\bf a)} $\tau=14.4\times 10^{-6}sec$

{\bf b)} $\tau=14.4\times 10^{-5}sec$

{\bf c)} $\tau=14.4\times 10^{-4}sec$

{\bf d)} $\tau=14.4\times 10^{-3}sec$

 {\bf FIG. 2 }  Momentum distribution after a readout
$Y=0$ has been detected. All parameters are as in fig. 1.

{\bf a)} $\tau=14.4\times 10^{-6}sec$

{\bf b)} $\tau=14.4\times 10^{-5}sec$

{\bf c)} $\tau=14.4\times 10^{-4}sec$

{\bf d)} $\tau=14.4\times 10^{-3}sec$


\begin{thebibliography}{100}
\bibitem{} E.T.Jaynes and F.Cummings, Proc.IEEE \textbf{51}, 89 (1963).
\bibitem{} H.I.Yoo, J.J.Sanchez-Mondragon and J.H.Eberly,
J.Phys.A:Math.Gen \textbf{14}, 1383 (1981).
\bibitem{} J.R.Kuklinski and J.Madajczyk, Phys.Rev.A \textbf{37}, 3175 (1988).
\bibitem{} S.M.Barnett, Opt.Commun. \textbf{61} 432 (1982); P.Zhou
and J.S.Peng, Phys.Rev.A \textbf{44} 3331 (1991)
\bibitem{} G.S.Agarwal, J.Opt.Soc.Am.B \textbf{2} 480 (1985).
\bibitem{} S.J.D.Phoenix and P.L.Knight, Ann.Phys.(N.Y.)
\textbf{186} 381 (1988).
\bibitem{} S.J.D.Phoenix and P.L.Knight, Phys.Rev.A \textbf{44} 6023 (1991).
\bibitem{} A.Ekert and P.L.Knight, Am.J.Phys. \textbf{63} 415 (1995).
\bibitem{} A.S.Soronsen and K.Molmer, Phys.Rev.Lett. \textbf{91}, 097905 (2003).
\bibitem{} J.I.Cirac, R.Blatt, A.S.Parkins, and P.Zoller, Phys.Rev.Lett. \textbf{70}, 762 (1993).
\bibitem{} D.M.Meekhof, C.Monroe, B.E.King, W.M.Itano, and D.J.Wineland, Phys.Rev.Lett. \textbf{76},  1796 (1996).
\bibitem{} W.J.Munro, Kao Nemoto, R.G.Beau Soleil, and T.P.Spiller, Phys.Rev.A \textbf{71}, 033819 (2005).
\bibitem{} W.G.Unruh, Phys.Rev.D \textbf{18}, 1764 (1978).
\bibitem{} G.J.Milburn and D.F.Walls, Phys.Rev.A \textbf{28}, 2065 (1980).
\bibitem{} V.Braginsky and F.I.Khalili, Zh.Eksp.Teor.Fiz. \textbf{78}, 1712 (1980).
\bibitem{} T.Sleator and M.Wilkens, Phys.Rev.A \textbf{48}, 3286 (1993).
\bibitem{} C.Adamas, M.Sigel, and J.Mlynek, Phys.Rep. \textbf{240}, 143 (1994).
\bibitem{} A.Kastberg, W.D.Philips, S.L.Rolston, R.J.C.Spreeuw, and P.S.Jessen, Phys.Rev.Lett. \textbf{74}, 1542 (1995).
\bibitem{} S.Yu, H.Rauch, and Y.Zhang, Phys.Rev.A \textbf{54}, 2585 (1995).
\bibitem{} R.L.de Matos Filho and W.Vogel, Phys.Rev.Lett. \textbf{76}, 608 (1996).
\bibitem{} J.R.Kuklinski and J.Madajczyk, Phys.Rev.A \textbf{48}, 3291 (1993).
\bibitem{} A.Auffeves, P.Maioli, T.Meunier, S.Gleyzes, G.Nogues, M.Brune, J.M.Raimond, and S.Haroche, Phys.Rev.Lett. \textbf{91}, 230405 (2003).
\bibitem{} C.C.Gerry, Phys.Rev.A \textbf{59}, 4095 (1999).
\bibitem{} B.Yurke and D.Stoler, Phys.Rev.Lett. \textbf{57}, 13 (1989).
\bibitem{} K.P.Marzlin and J.Audertsch, Phys.Rev.A \textbf{53}, 1004 (1996).
\bibitem{} W.P.Schleish, \textit{Quantum optics in phase space}, (Springer,VCH,2001).
\bibitem{} G.J.Milburn and D.F.Walls, Phys.Rev.A \textbf{28}, 2065 (1983).
\bibitem{} N.Imoto, S.Watkins, and Y.Sasaki, Opt.commun. \textbf{61}, 159 (1987).
\bibitem{} M.Brune, S.Haroche, V.Lefevre, J.M.Raimond, and N.Zagury, Phys.Rev.Lett. \textbf{65}, 976 (1990).
\bibitem{} V.B.Braginsky and S.P.Vyatohanin, Phys.Lett.A \textbf{132}, 206 (1988).
\bibitem{} P.Storey, M.Collett and D.Walls, Phys.Rev.Lett. \textbf{68}, 472 (1992).
\bibitem{} M.A.M.Marte and Zoller, Appl.Phys.B \textbf{54}, 477 (1992).
\end{thebibliography}
\end{document}